\documentstyle[floats,aps,aipbook,psfig]{revtex}
\righthead{Norris {\it et al.}}
\lefthead{Time Dilation of Intervals}
\begin{document}
\title{Test for Time Dilation of Intervals Between Pulse Structures in GRBs}
\author{J.P. Norris,$^1$ J.T. Bonnell,$^2$ 
   R.J. Nemiroff,$^3$ and J.D. Scargle$^4$}
\address{$^1$NASA/Goddard Space Flight Center, Greenbelt, MD 20771\\
         $^2$Universities Space Research Association\\
         $^3$George Mason University, Fairfax, VA 22030\\
         $^4$NASA/Ames Research Center, Moffett Field, CA 94035}

\maketitle

\begin{abstract}
If $\gamma$-ray bursts are at cosmological distances, then not
only their constituent pulses but also the intervals between pulses should be
time-dilated.  Unlike time-dilation measures of pulse emission, intervals
would appear to require negligible correction for redshift of narrower
temporal structure from higher energy into the band of observation.   However,
stretching of pulse intervals is inherently difficult to measure without
incurring a timescale-dependent bias since, as time profiles are stretched,
more structure can appear near the limit of resolution.  This problem is
compounded in dimmer bursts because identification of significant structures
becomes more problematic.  We attempt to minimize brightness bias by
equalizing signal-to-noise (s/n) level of all bursts.   We analyze
wavelet-denoised burst profiles binned to several resolutions, identifying 
significant fluctuations between pulse structures and interjacent valleys.  
When bursts are ranked by peak flux, an interval time-dilation signature 
is evident, but its magnitude and significance are dependent upon
temporal resolution and s/n level.
\end{abstract}

\section*{Expected Systematics}

It might naively be thought that time dilation of intervals between peaks or
pulses in $\gamma$-ray bursts (GRB) should be free of the energy-dependent 
effects that plague measures of time dilation of temporal structures
\cite{{JPN1},{JPN2}}.  However, at least two effects are expected to give rise
to systematic biases that make attempts to determine the actual measure of
time dilation of pulse intervals difficult:

\paragraph*{(1) Structure appearing at limit of temporal resolution.}
For the present purpose, define ``interval between pulses'' to be the interval
between two discernible peaks of emission, desired significance being
adjustable.  The average width of GRB pulses in long ($T_{\rm 90}$ $>$ 2 s) 
bursts is $\sim$ 100--500 ms, dependent on energy band.  However, there is a 
large dispersion in pulse width \cite{JPN3}.  Since the timescale for
intervals between pulses is also of this order, there is often a high degree 
of pulse overlap.  Consequently, for the 64-ms resolution data which we employ,
time dilating a burst profile by a factor $S$ $\sim$ 2 will result in the 
appearance of newly resolved structure at the shortest resolved timescale.  
Some intervals between peaks which were not resolved in the unstretched burst
will then become discernible, with the result that some intervals in the 
original profile divide into two shorter ones.  The average pulse interval in 
a stretched burst will therefore not be $S$ times longer than in the 
unstretched burst, but somewhat less than $S$.

\paragraph*{(2) Redshift of narrower pulses from higher energy 
(deeper valleys).}
This is the same effect which diminishes the observed measure of time dilation
in pulse widths:  Since pulses are narrower at higher energy, the dimmer
bursts -- presumably suffering more redshift -- will have narrower pulses
redshifted into the band of observation.  The effect on pulse-interval
measures is that valleys between pulses will be deeper and more significant 
in the redshifted bursts since there will be less pulse overlap.  This effect 
will result in additional (otherwise time-dilated) intervals being bifurcated, 
and therefore shortened.  Also, some new valleys will appear that were not 
present in the non-redshifted burst profile.  

\paragraph*{Measures of pulse-interval dilation.}
A time-dilation measure for intervals which appears relatively unbiased is
the average (or median) interval between pulses or peaks, per burst (see also 
the definitions in ref \cite{Neubauer}, these proceedings).  Alternatively, 
all intervals found within a given brightness group might be weighted equally 
\cite{Davis}, but this would tend to weight longer bursts more heavily.  A 
measure like event duration is the interval between first and last significant 
peaks.  A more complex formulation might take into account the significance 
(e.g., depth of interjacent valley) of an interval.  How such definitions are
to be corrected (assuming cosmological hypothesis) for redshift and resolution 
effects should be estimated by performing simulations.  In this paper we report 
results only for a test of the time-dilation effect between pulse intervals, 
and leave the understanding of corrections for a more detailed study.

\section*{Procedure}

\paragraph*{Data preparation.} 

BATSE bursts in the 3B catalog with measured $T_{\rm 90}$ $>$ 2 s \cite{JTB}
above a peak-intensity threshold form the sample.  The threshold is either 
2400 or 1400 counts s$^{-1}$, with the sample divided into 5 or 6 groups
($\sim$ 85 bursts per group), respectively, according to BATSE 3B peak flux
(256-ms timescale).  BATSE DISCSC data (64-ms resolution) summed over channels 
1--4 ($>$ 25 keV) is used; quadratic (infrequently, higher order) backgrounds 
are fitted and subtracted.  Burst profiles are prepared by rendering their 
signal-to-noise (s/n) levels equal to that of the burst with the lowest peak 
intensity in the sample, according to a procedure discussed in ref \cite{JPN1}. 
This step renders variances and peak intensities approximately equal.  The 
prepared profiles are then run through a Haar wavelet-denoiser to remove 
insignificant ($<$ 2-$\sigma$) fluctuations on all timescales.  Without 
denoising, identification of some valleys would often be compromised by
insignificant fluctuations.

\paragraph*{Interval identification.}
The profiles are searched for occurrences of two peaks separated by a valley,
requiring a significance of the intensity difference of at least 4 $\sigma$ 
between the lower peak and the valley.  By requiring a highly significant 
interval, we are essentially identifying intervals between major pulse 
structures, rather than individual pulses, thereby (hopefully) ameliorating
some systematic effects described in the previous section.  The interval
search is performed for the prepared profiles binned to 64-ms, 128-ms, 256-ms,
and 512-ms resolutions.  Each binning timescale was analyzed separately.

\begin{table}
\caption{Interval Time-Dilation Factors vs 3B Peak Flux, Resolution}
\label{table1}
\begin{tabular}{llllll}
&
\multispan{5} \hfil{Peak Flux (ph cm$^{-2}$ s$^{-1}$)}
  \tablenote{Lower peak-flux boundary for 5 brightness groups; 
  boundary for brightest group: 4.60 ph cm$^{-2}$ s$^{-1}$.  Values
  in parentheses are probabilities for stretch factor of unity.}\hfil\\
        & {2.10}       & {1.33}       & {0.93}       & {0.65}        & {0.32}\\
\tableline
&
\multispan{5} \hfil{\bf Threshold: 1400 cts s$^{-1}$}\hfil\\
 64 ms  & 1.55 (0.38)  & 2.85 (0.035) & 1.25 (0.87)  & 3.25 (0.018) 
        & 2.30 (0.054)\\
128 ms  & 1.65 (0.085) & 1.72 (0.072) & 1.75 (0.13)  & 2.35 (0.0008) 
        & 2.42 (0.004)\\
256 ms  & 1.28 (0.36)  & 1.35 (0.59)  & 1.25 (0.77)  & 1.98 (0.013) 
        & 1.50 (0.14)\\
512 ms  & 1.15 (0.45)  & 1.35 (0.25)  & 1.60 (0.093) & 2.18 (0.016) 
        & 2.38 (0.0013)\\

\\
&
\multispan{5} \hfil{\bf Threshold: 2400 cts s$^{-1}$}\hfil\\
 64 ms  & 0.85 (0.67)  & 1.18 (0.32)  & 1.28 (0.20)  & 1.72 (0.029)\\
128 ms  & 1.30 (0.22)  & 1.75 (0.032) & 1.40 (0.29)  & 2.15 (0.008)\\
256 ms  & 1.05 (0.60)  & 1.55 (0.037) & 1.40 (0.10)  & 2.20 (0.0016)\\
512 ms  & 1.05 (0.82)  & 1.30 (0.21)  & 1.22 (0.62)  & 2.20 (0.0002)\\
\end{tabular}
\end{table}

\section*{Results}

We adopt the first measure of interval time-dilation described above, the
median interval per burst.  We then form distributions of median intervals for
each brightness group.  By stretching the distribution of intervals for the
brightest group on a grid of trial time-dilation factors and performing a
Kolmogorov-Smirnov (K-S) test for degree of agreement between the interval
distribution of the brightest group and those of the five dimmer groups, we
estimate observed time-dilation factors and associated errors.

For each binning timescale and the two peak intensity thresholds, 
Table \ref{table1} lists the measured interval time-dilation factors (TDF), and
the probability (in parentheses) of agreement given a stretch factor of unity, 
of the five dimmer groups relative to the brightest group.  More significant 
determinations result more often for the higher peak intensity threshold, 
presumably because a higher s/n level is realized in the noise equalization 
procedure.  However, the higher threshold necessarily cannot examine the 
dimmer bursts.  For all timescales a trend is evident of longer median 
intervals towards lower peak flux.  

Significances of disagreement between the interval distributions of dimmest 
(or second dimmest) and brightest groups range from ${\sim}$ 2-${\sigma}$ to 
3.5-${\sigma}$, with longer timescales tending to be more significant.
A partial explanation for this must be that for coarser binning (higher
counts per bin), a larger number of bursts survive to contribute to the 
distribution:  the numbers of occurrences of bursts with 2 or more peaks with 
a $>$ 4-${\sigma}$ valley in between increases as the timescale increases.  
For 1400 counts s$^{-1}$ threshold, the number of such occurrences per group 
increases from ${\sim}$ 24 (64 ms) to 38 (256-512 ms); for 2400 counts s$^{-1}$ 
threshold, the number of contributing bursts is approximately constant with
timescale, ${\sim}$ 50 occurrences.

Two examples of the trend are illustrated in Figures 1 and 2.  
The first case is for the 1400 counts s$^{-1}$ threshold on the longest 
timescale analyzed, 512 ms; the second case shows results for 
the 2400 counts s$^{-1}$ threshold for 128 ms resolution.  Both figures 
illustrate the more conservative result (in terms of significance) for their 
respective timescales obtained for the bright group relative to dimmest 
(or second dimmest) group, as can be seen by comparing probabilities for unity 
stretch factor for the two thresholds in Table 1.

\section*{Conclusions}

When bursts are grouped by BATSE peak flux, we find a relative
time-dilation effect for intervals between pulse structures, at a significance
level of $\sim$ 2.5 $\sigma$, between brightest and dimmest / next to dimmest
burst groups.  This {\it observed} time-dilation factor is of order 2. 
Actual time-dilation factors would probably be somewhat larger:  Two effects --
appearance of new structure at the limit of resolution as bursts are
stretched, and narrowing of pulses with higher energy, thus better defining 
valleys between pulse structures -- probably result in the observed
time-dilation factor being smaller than the actual value.  As this is an
exploratory study in need of robust simulations to calibrate these effects,
we conclude that this result tentatively and qualitatively confirms the result
of Davis \cite{Davis}, in which intervals between pulse structures were 
measured using a pulse-fitting approach.

\begin{figure*}
\leavevmode
\psfig{file=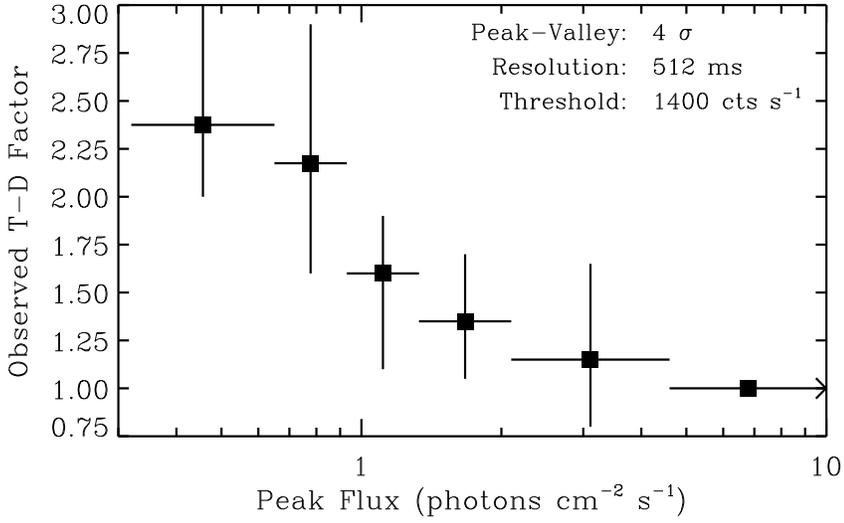,height=3.0in,width=5.0in}
\caption{
Observed interval time-dilation factor vs. BATSE 3B peak flux, for
1400 counts s$^{-1}$ threshold for profiles rendered to 512 ms resolution. 
Central values and 1-${\sigma}$ uncertainties determined via K-S test, by 
stretching distributions of intervals for bright burst group and comparing 
with distributions of dimmer groups.}
\end{figure*}

\begin{figure*}
\leavevmode
\psfig{file=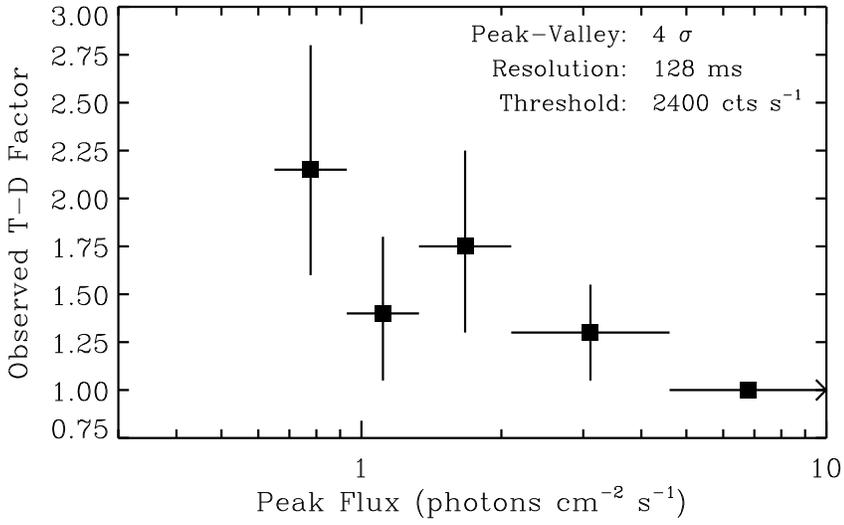,height=3.0in,width=5.0in}
\caption{
Observed interval time-dilation factor vs. BATSE 3B peak flux, for
2400 counts s$^{-1}$ threshold for profiles rendered to 128 ms resolution.}
\end{figure*}

\end{document}